\documentclass[aps,prl,twocolumn,groupedaddress,superscriptaddress,nobibnotes]{revtex4}
\usepackage{amsmath}
\usepackage{amssymb}
\usepackage{graphicx}
\usepackage{dcolumn}
\usepackage{bm}
\usepackage{color}
\usepackage[normalem]{ulem}

\begin{document}

\title{Relocation of the topological surface state of Bi$_{2}$Se$_{3}$ beneath the surface by Ag intercalation}

\author{M. Ye}
\altaffiliation{Present address: Hiroshima Synchrotron Radiation Center, Hiroshima
University, Japan.}
\affiliation{Graduate School of Science, Hiroshima University, 1-3-1
Kagamiyama, Higashi-Hiroshima, 739-8526, Japan}

\author{S. V. Eremeev}
\affiliation{%
    Institute of Strength Physics and Materials Science, 634021, Tomsk,
    Russia}
\affiliation{%
    Tomsk State University, 634050, Tomsk, Russia}

\author{K. Kuroda}
\affiliation{Graduate School of Science, Hiroshima University, 1-3-1
Kagamiyama, Higashi-Hiroshima, 739-8526, Japan}

\author{M. Nakatake}
\affiliation{Hiroshima Synchrotron Radiation Center, Hiroshima
University, 2-313 Kagamiyama,  Higashi-Hiroshima, 739-0046, Japan}

\author{S. Kim}
\affiliation{Graduate School of Science, Hiroshima University, 1-3-1
Kagamiyama, Higashi-Hiroshima, 739-8526, Japan}

\author{Y. Yamada}
\affiliation{Graduate School of Science, Hiroshima University, 1-3-1
Kagamiyama, Higashi-Hiroshima, 739-8526, Japan}

\author{E.~E.~Krasovskii}
\affiliation{%
Departamento de F\'{\i}sica de Materiales UPV/EHU and Centro de
F\'{\i}sica de Materiales CFM and Centro Mixto CSIC-UPV/EHU, 20080
San Sebasti\'an/Donostia, Basque Country, Spain
\\
}
\affiliation{%
Donostia International Physics Center (DIPC),
             20018 San Sebasti\'an/Donostia, Basque Country,
             Spain\\
}
\affiliation{%
IKERBASQUE, Basque Foundation for Science, 48011 Bilbao, Spain\\
}

\author{E.~V.~Chulkov}
\affiliation{%
Departamento de F\'{\i}sica de Materiales UPV/EHU and Centro de
F\'{\i}sica de Materiales CFM and Centro Mixto CSIC-UPV/EHU, 20080
San Sebasti\'an/Donostia, Basque Country, Spain
\\
}
\affiliation{%
Donostia International Physics Center (DIPC),
             20018 San Sebasti\'an/Donostia, Basque Country,
             Spain\\
}

\author{M. Arita}
\affiliation{Hiroshima Synchrotron Radiation Center, Hiroshima
University, 2-313 Kagamiyama,  Higashi-Hiroshima, 739-0046, Japan}

\author{H. Miyahara}
\affiliation{Graduate School of
Science, Hiroshima University, 1-3-1 Kagamiyama, Higashi-Hiroshima,
739-8526, Japan}

\author{T. Maegawa}
\affiliation{Graduate School of
Science, Hiroshima University, 1-3-1 Kagamiyama, Higashi-Hiroshima,
739-8526, Japan}

\author{K. Okamoto}
\affiliation{Graduate School of
Science, Hiroshima University, 1-3-1 Kagamiyama, Higashi-Hiroshima,
739-8526, Japan}

\author{K. Miyamoto}
\affiliation{Hiroshima Synchrotron Radiation Center, Hiroshima
University, 2-313 Kagamiyama,  Higashi-Hiroshima, 739-0046, Japan}

\author{T. Okuda}
\affiliation{Hiroshima Synchrotron Radiation Center, Hiroshima
University, 2-313 Kagamiyama,  Higashi-Hiroshima, 739-0046, Japan}

\author{K. Shimada}
\affiliation{Hiroshima Synchrotron Radiation Center, Hiroshima
University, 2-313 Kagamiyama,  Higashi-Hiroshima, 739-0046, Japan}

\author{H. Namatame}
\affiliation{Hiroshima Synchrotron Radiation Center, Hiroshima
University, 2-313 Kagamiyama,  Higashi-Hiroshima, 739-0046, Japan}

\author{M. Taniguchi}
\affiliation{Graduate School of Science, Hiroshima University, 1-3-1
Kagamiyama, Higashi-Hiroshima, 739-8526, Japan}
\affiliation{Hiroshima Synchrotron Radiation Center, Hiroshima
University, 2-313 Kagamiyama,  Higashi-Hiroshima, 739-0046, Japan}

\author{Y. Ueda}
\affiliation{Kure National College of Technology, Agaminami 2-2-11,
Kure 737-8506, Japan}

\author{A. Kimura}
\email{akiok@hiroshima-u.ac.jp} \affiliation{Graduate School of
Science, Hiroshima University, 1-3-1 Kagamiyama, Higashi-Hiroshima,
739-8526, Japan}

\date{\today}

\begin{abstract}
We studied the Ag-intercalated 3D topological insulator
Bi$_{2}$Se$_{3}$ by scanning tunneling microscopy/spectroscopy and
angle-resolved photoemission spectroscopy, combined with a first
principles calculations. We demonstrate that silver atoms deposited
on the surface of Bi$_{2}$Se$_{3}$ are intercalated between the
quintuple layer (QL) units of the crystal, causing a expansion of
the van der Waals gaps and the detachment of topmost QLs from the
bulk crystal. This leads to a relocation (in the real space) of the
the topological state beneath the detached quintuple layers,
accompanied by the emergence of parabolic and ``M-shaped'' trivial
bands localized above the relocated topological states. These novel
findings open a pathway to the engineering of Dirac fermions
shielded from the \emph{ambient} contamination and may facilitate
the realization of fault-tolerant quantum devices.
\end{abstract}

\maketitle

\begin{figure}
\includegraphics[width=\columnwidth]{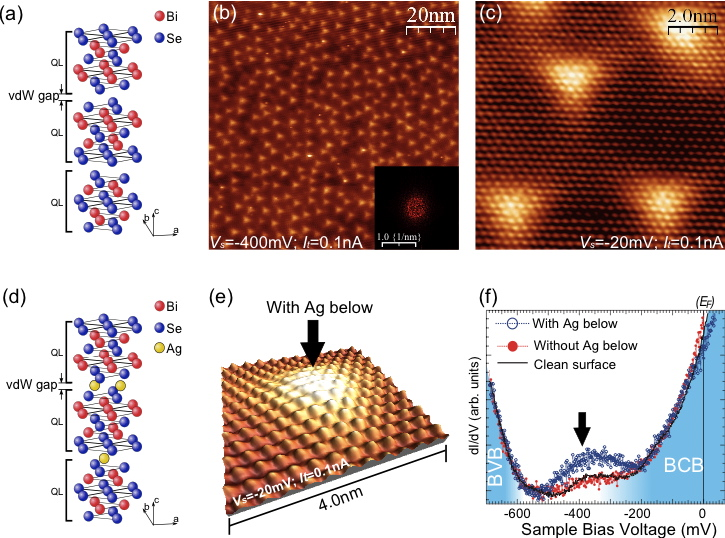}
 \caption{ \label{fig:epsart1} (Color online) (a) Crystal structure of Bi$_{2}$Se$_{3}$. (b) Topography STM image of Ag deposited Bi$_{2}$Se$_{3}$ surface with triangular-shaped corrugations in 100 nm ~ 100 nm area. Inset: Fourier transformed image of (b). (c) Atomically resolved STM image of Ag deposited surface. (d) Schematic model of Ag intercalated Bi$_{2}$Se$_{3}$. (e) 3-dimensional illustration of the magnified image of one of the convexities with Ag intercalated below (marked by an arrow). (f) Spatially resolved tunneling spectra taken at the areas with and without (-40mV shifted) Ag atoms intercalated underneath, compared with the spectrum of clean surface shifted by -135 mV.}
\end{figure}

A novel class of quantum materials, called topological insulators (TIs) \cite{Zhang09NP, Hasan2010RMP}, has provoked much research interest. A number of materials that hold nontrivial spin-polarized metallic surface states have been intensively studied, such as Bi$_{1-x}$Sb$_{x}$ \cite{Hsieh2009Science, Roushan2009Nature, Nishide2010PRB}, Bi$_{2}$Te$_{3}$ \cite{ChenYL2009Science, ZhangT2009PRL}, Bi$_{2}$Se$_{3}$ \cite{XiaY2009NP, Kuroda_PRL10a, ZhangY2010NP}, and thallium-based TIs \cite{EremeevJETPL_Th_10, Kuroda_PRL10, Yan_EPL10, Lin_PRL10, Sato_PRL10, Chen_PRL10}, among which Bi$_{2}$Se$_{3}$ is one of the most promising candidates for potential applications in ultra-low power consumption quantum devices that can work stably at room temperature due to a sufficiently large bulk energy gap \cite{XiaY2009NP}. Owing to the time-reversal symmetry, topological surface states are protected from backscattering in the presence of a weak perturbation, which is important for the realization of dissipationless spin transport in novel quantum devices.

However, such surface state is not protected from scattering by arbitrary angles, so the shielding of this state from impurities is an important problem. In this letter, we demonstrate that silver atoms deposited on the surface of Bi$_{2}$Se$_{3}$ are intercalated between the quintuple layer (QL) units of the crystal, causing their detachment from the substrate. This leads to a relocation in the real space of the topological surface state beneath the detached quintuple layers, accompanied by the appearance of trivial two-dimensional (2D) states. These novel findings open a pathway to the engineering of Dirac fermions shielded from the ambient contamination and may facilitate the realization of fault-tolerant quantum devices \cite{Kitaev2003AP}.

The single crystalline of Bi$_{2}$Se$_{3}$ used in the present work was grown by a standard Bridgman method, which can be found elsewhere\cite{Kuroda_PRL10}. The deposition of Ag atoms were conducted by setting a small amount of Ag wire in a conventional filament basket made by tungsten wire. The scanning tunneling microscopy/spectroscopy (STM/STS) measurements were performed with a commercial low-temperature STM (LT-STM, Omicron). The sample was cooled to 78 K by liquid nitrogen during STM/STS measurement. The STM image presented in this work were all taken in a constant current mode with an electrochemically etched tungsten tip. The STS data and differential conductance map were collected with a standard lock-in technique working at a frequency of 1.0 kHz and modulation amplitude of 10 mV. The ARPES experiments were performed at circular polarized undulator beamline (BL-9A and BL-9B) of Hiroshima Synchrotron Radiation Center (HSRC) equipped with a hemispherical photoelectron analyzers (VG-SCIENTA R4000). The energy resolution was set at 8 - 15 meV during the experiments.

The crystal structure of Bi$_{2}$Se$_{3}$ is built of quintuple units consisting of hexagonal atomic layers in the order of Se-Bi-Se-Bi-Se, as depicted in Fig. 1(a). Hence, the clean surface obtained by in situ cleavage in ultrahigh vacuum (UHV) exposes the topmost Se-layer of the QL. Figure 1(b) shows a typical topographic STM image of the Bi$_{2}$Se$_{3}$ surface measured at 78 K after the deposition of Ag at room temperature. The surface is seen to be dominated by triangular-shaped bright spots. A further analysis of the topographic image by Fourier transformation (inset of Fig. 1(b)) reveals a circular shaped intensity distribution in the center of the transformed image, whose radius 0.176 nm$^{-1}$ does not depend on the sample bias voltage. This means that the triangular-shaped convexities are uniformly distributed over the surface with a separation of around 5.6 nm. The bright feature observed on the surface by STM is almost independent of the sample bias voltage, indicating that the different brightness of the image results from the surface morphology rather than from the electronic structure. The atomically resolved STM image of a smaller region is shown in Fig. 1(c). The six-fold symmetry arrangement of Se lattice on the topmost surface is clearly seen. More important, the Se layer appears as a periodic lattice continuously spreading over the surface with triangular-shaped convexities, indicating that all the Ag atoms are beneath the topmost layer. It is reasonable to suppose that Ag atoms are intercalated in the van der Waals gaps between QLs of Bi$_{2}$Se$_{3}$ crystal (Fig. 1(d)) due to the weak bonding. This point will be thoroughly discussed later. Figure 1(e) shows a 3D image of a small-area scan containing one of the triangular-shaped regions in Fig. 1(c). The surface Se layer is seen to be slightly protruded due to the intercalated Ag atoms underneath. The bright region in this figure, indicated by a black arrow, corresponds to the area with buried Ag atoms.

To further explore the electronic states of the Ag-intercalated Bi$_{2}$Se$_{3}$ surface, we have measured the spatially resolved differential conductance spectra of the corrugated surface. They are known to reflect to a good approximation the local density of states (LDOS) of the probed surface area \cite{Tersoff1985PRB}. The spectra shown in Fig. 1(f) are taken in the region with (blue) and without (red) Ag underneath, respectively, corresponding to the bright and dark areas in the topographic image. For comparison, also the spectrum taken on the clean Bi$_{2}$Se$_{3}$ surface is shown in Fig. 1(f). The steep edges below and above the energy gap in the three $dI/dV$ spectra coincide to within an energy shift. The shift of $\sim$ 135 mV between the spectrum of the area with Ag and the clean surface is ascribed to the electron doping due to Ag intercalation, while the small shift of $\sim$ 40 mV between the spectra of the areas with and without Ag atoms is caused by the spatially local energy shift of the electron states relative to the Fermi level due to the influence of the intercalated Ag atoms.

Note that the structure of the $dI/dV$ curve of the Ag intercalated surface (Fig. 1(f)) in the bulk energy gap is quite different from that of the clean surface. In the region with Ag (blue), the spectrum has a pronounced W-like shape with an intensity enhancement around -400 mV marked by a black arrow, in contrast to the V-shaped spectrum of the clean surface (black solid line). The spatial distribution of the $dI/dV$ intensity is strongly inhomogeneous too: in the valleys it is much weaker than at the convexities.

In order to further investigate the changes of the electronic states both at the surface and at the interface where Ag atoms are intercalated, we have performed a probing-depth-sensitive angle-resolved photoelectron spectroscopy (ARPES) measurement utilizing a tunable low-energy synchrotron radiation. Figure 2(a) shows the ARPES images of the evolution of the surface band structure upon Ag deposition time taken with a photon energy of $h\nu$ = 21 eV. In the case of the clean surface, the cone-shaped dispersion is clearly seen with the Dirac point located at the binding energy ($E_{B}$) of 300 meV. Simultaneously, the bulk conduction band can be observed near the Fermi level ($E_{F}$) with the band bottom located at $E_{B}$ = 120 meV. After a small amount of Ag deposited on the surface (0.5 minute), both the bulk and the surface bands shift by 60 meV. The energy shift stabilizes at 80 meV after 3 minutes of Ag deposition. Besides the energy shift of the bulk energy bands, it is important that after the Ag deposition there appear two surface-state bands, which overlap with the bulk conduction band near the conduction band minimum (CBM), as indicated by red dashed box in the rightmost graph of Fig. 2(a). An expanded image near the CBM is shown in Fig. 2(b), where the peak positions fitted from energy distribution curves (EDC) are plotted by red square marks. The band bottoms of these two bands are determined as 79 meV and 186 meV by fitting the plots with two parabolic curves (black dashed lines), respectively.

\begin{figure}
\includegraphics[width=\columnwidth]{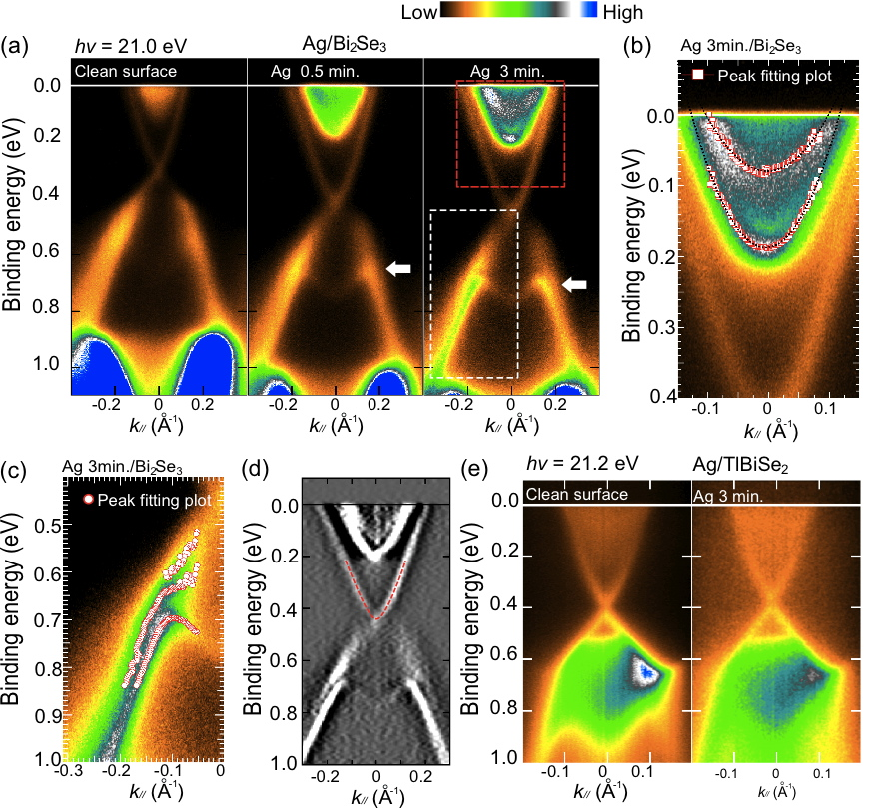}
 \caption{ \label{fig:epsart2} (Color online) (a) Band structures of Bi$_{2}$Se$_{3}$ with the increase of the Ag deposition time from 0.5 to 3 minutes measured by ARPES with photon energy of 21.0 eV. (b)and (c), expanded image of the red and white dashed boxes in (a) with superimposed  EDC peak dispersion plots (red squares and circles). (d) 2nd derivative ARPES image of Bi$_{2}$Se$_{3}$ after Ag deposition measured by $h\nu$ of 21.0 eV. (e) Band structures of TlBiSe$_{2}$ with 3 minutes Ag deposition.}
\end{figure}

Another important observation is that as a result of the Ag deposition a new M-shaped feature gradually develops in the local gap of the valence band below the Dirac cone, at $E_{B}$ $\sim$ 700 meV, see the white arrow in the rightmost graph of Fig. 2(a). It becomes more pronounced as the Ag amount increases. Notably, the spectral weight of the background does not appreciably increase even after 5 minutes of Ag deposition, which means that the Ag-deposited Bi$_{2}$Se$_{3}$ surface preserves the periodicity. We have been able to resolve the fine structure of the M-shaped band: in Fig. 2(c) we show a magnified view of the region indicated by the white dashed line box in Fig. 2(a). The peak fitting analysis of the EDCs and MDCs (momentum distribution curves) reveals several dispersion branches in this region. Interestingly, although our samples measured by ARPES prior to Ag deposition show quite different carrier densities, with the Dirac point located at different binding energies, after Ag deposition the corresponding electronic states appear at the same binding energy in different samples (not shown). It should be noted that after Ag deposition, the electronic state in the bulk energy gap appears to behave as a trivial surface state rather than as a Dirac state. Fig. 2(d) shows the 2nd derivative ARPES data acquired after 5 minutes Ag deposition, where in the bulk energy gap, instead of a gapless state with a Dirac point, we find a new parabolic energy band with the minimum at $E_{B}$ = 410 meV, as indicated by the red dashed line.

Similar 2D states split off from the bulk conduction and bulk valence bands have been reported by several authors, where their origins have been ascribed to a band bending effect induced by adsorbed atoms or molecules on the surface \cite{Bianchi10NC, Bianchi11PRL, King11PRL, Benia11}. Nevertheless, no direct evidence on the band bending effect has been provided in their studies. In our Ag deposition case, on the other hand, the 2D states are generated as a result of the Ag intercalation as observed by STM (see Fig.1). In order to further confirm the intercalation effect rather than the band bending as a possible origin of of the trivial 2D states, we have performed a Ag-deposition experiment for the rather three-dimensional material TlBiSe$_{2}$ \cite{Kuroda_PRL10},  where no vdW gap exists and thus the intercalation should hardly take place. As shown in Fig.2(e), the Ag deposition leads only the increased background but  does not give rise to any trivial 2D states. This result strengthen our conclusion that the trivial 2D states are originated from the Ag-intercalation into the vdW gap of Bi$_{2}$Se$_{3}$.

In order to increase the bulk sensitivity of the measurements we have performed the ARPES experiment with the photon energy of 7.95 eV. Figure 3(a) compares the energy-momentum intensity distribution of Bi$_{2}$Se$_{3}$ acquired after 3 minutes Ag deposition with two different excitation energies. After the measurement with the photon energy of 21.0 eV, left panel of Fig. 3(a), the photon energy was immediately lowered to 7.95 eV, and the measurement has restarted within less than 5 minutes. The right panel of Fig. 3(a) shows the data measured with 7.95 eV photons, where the contrast of the bulk conduction band is strongly enhanced. Simultaneously the two parabolic bands near $E_{F}$ observed for $h\nu$ = 21.0 eV are strongly suppressed. In the bulk-sensitive measurement at $h\nu$ =7.95 eV we observe the electronic structure very close to that of the clean Bi$_{2}$Se$_{3}$ surface: the Dirac point is clearly seen 0.08 eV above the band minimum observed at $h\nu$ =21.0 eV.  This is illustrated by the k$_{||}$ dispersion of intensity maxima obtained by the MDC fitting procedure, see Fig. 3(b).

\begin{figure}
\includegraphics[width=\columnwidth]{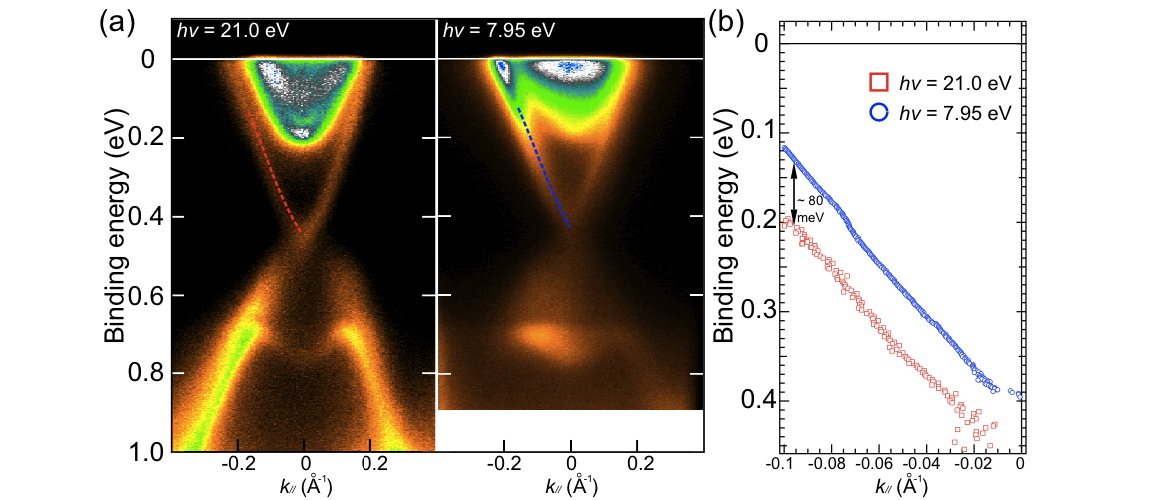}
 \caption{\label{fig:epsart3} (Color online) (a) Band structure of Bi$_{2}$Se$_{3}$ after 3 minutes Ag deposition measured with photon energies of 21.0 eV (left) and 7.95 eV (right). (b) Dependence of the binding energy location of the EDC peak on $k_{||}$ for $h\nu$ = 21.0 eV (red dashed line in (a)) and for $h\nu$ = 7.95 eV (blue dashed line in (a)) shows a large energy difference of 80 meV between the bands observed at different $h\nu$}
\end{figure}

Thus, at the surface of the Ag-intercalated Bi$_{2}$Se$_{3}$ there exist both trivial 2D states and a topological subsurface state. To gain further insight into the novel electronic states, we have performed a first-principles band structure calculation using the VASP code \cite{VASP1, VASP2}. The interaction between the ion cores and valence electrons was described by the projector augmented-wave method \cite{PAW1, PAW2}. The generalized gradient approximation \cite{PBE} was used to describe the exchange correlation energy. The Hamiltonian contained the scalar-relativistic corrections, and the spin-orbit coupling was taken into account by the second variation method \cite{Koelling}. The calculated band structure of an ideal Bi$_{2}$Se$_{3}$ is shown in Fig. 4(b). The topological surface state with a cone-shaped dispersion (red solid line) is well reproduced. The analysis of the charge density distribution profile (Fig. 4(c)) reveals that the Dirac state is mostly localized in the 1-st QL and almost completely decays within the two outermost QLs.

\begin{figure}
\includegraphics[width=\columnwidth]{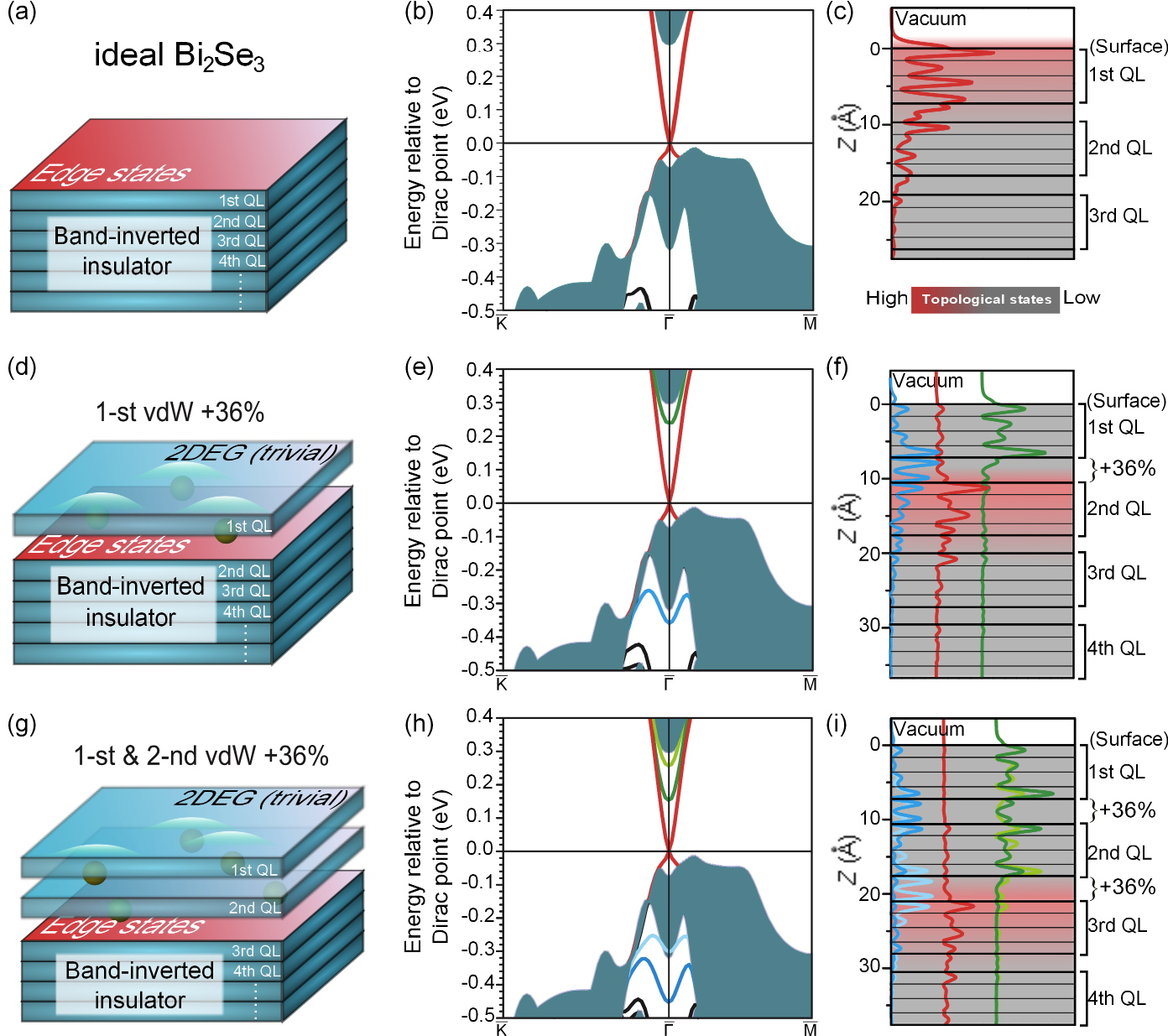}
 \caption{\label{fig:epsart5}(Color online)(a) Schematic model of ideal
Bi$_{2}$Se$_{3}$. (b) Calculated electronic structures of ideal
Bi$_{2}$Se$_{3}$. (c) Charge density distribution profile of the
Dirac cone in ideal Bi$_{2}$Se$_{3}$. (d) Schematic model for the
expansion of the first vdW gap caused by Ag intercalation. (e)
Electronic structures calculated for the model in (d), with the
first vdW gap expanded by 36\%. (f) Localization of the calculated
electronic states in the corresponding colors used in (e). (g)-(i)
Same calculation with both the first and second vdW gaps expanded.}
\end{figure}

We now consider the case of Ag intercalation. The main problem for the theoretical modeling comes from the unknown geometry of the intercalated Ag atoms. Let us tentatively assume that Ag atoms form a 2D monolayer island in the van der Waals (vdW) gap between the 1st and 2nd QL of Bi$_{2}$Se$_{3}$. According to our theoretical geometry optimization, the first vdW gap expands by 36\% when a full Ag monolayer is inserted. This is schematically shown in Fig. 4(d). The calculated electronic structure of a hypothetical system with the vdW gap increased by 36\% is shown in Fig. 4(e): a parabolic surface-state band (of Bi $p_{z}$ character) is seen to split from the bottom of bulk conduction band (green). It should be noted that the energy position of the parabolic band strongly depends on the expansion of the vdW gap (also see the Supplementary Information). Because after the Ag intercalation we observe a rather inhomogeneous surface in the STM experiment, the expansion of the vdW gap is, apparently, spatially inhomogeneous too, which explains the broadening of the parabolic bands observed in our ARPES experiments. This may also be the reason for the discrepancy in the energy position of the parabolic band relative to the bulk conduction band between the theory and the experiment.

The most interesting feature can be inferred from the spatial localization of the Dirac state as shown in Fig. 4(f) (the red solid line). Unlike the Dirac surface state localized at the ideal surface of the 3D topological insulator, the topological state (red) in the case of the 36\% expansion of the first vdW gap is buried 10-15 $\rm \AA$ below the surface, {\it i.e.}, just below the expanded vdW gap. At the same time, the detached QL behaves as a trivial insulator. As can be seen in Fig. 4(f) the parabolic band (green solid line) is confined inside of the detached QL on the top of the crystal, which explains why the surface sensitive ARPES measurements at 21.0 eV mainly probe the new parabolic band (green solid line in Figs. 4(e) and (f)), while the bulk sensitive measurement with 7.95 eV photon energy can detect the Dirac cone below the surface (red solid line in Fig. 4(f)). Apart from the parabolic band split from the bottom of the bulk conduction band, the calculations correctly reproduce the M-shaped band split from the top of the local gap in the valence band, blue line in Figs. 4(e) and (f). These surface states are localized predominantly on the Se atoms bordering the expanded vdW gap and have $p_{z}$ character. Two-dimensional states with similar dispersion -- parabolic in the conduction band and M-shaped in the valence band -- appear also in our calculation for a free-standing QL of Bi$_{2}$Se$_{3}$ (see Supplementary Information), in agreement with the earlier calculation by Zhang {\it et al.} \cite{ZhangY2010NP}. Also the charge distribution in the parabolic band is very similar to that in the detached layer (green lines in Fig. 4(f)): the charge is localized at Bi atoms of the free-standing QL. At the same time, the charge distribution in the M-shaped bands is different from the case of a free-standing QL, where the charge is localized at the outer Se layers of the slab, symmetrically with respect to its center. In the case of the detached QL the charge transfers from the outer surface to the inner surfaces of the expanded vdW gap.

We further consider a more complicated situation of the Ag intercalation as schematically described in Fig. 4(g), where Ag atoms are intercalated not only in the first vdW gap but also in the second vdW gap between the 2nd and 3rd QLs, so that the two top QLs are detached from the bulk crystal. We again assume an expansion of 36\% for both the first and the second vdW gaps. The calculated electronic structure for this system is shown in Fig. 4(h). Two parabolic bands (dark and light green) are seen to be split from the bottom of the bulk conduction band, and two M-shaped bands (dark and light blue) appear in the valence band gap. These trivial surface states are observed in our ARPES experiment (see Figs.2(a)-(c)). Further analysis of the localization of these electronic states, as shown in Fig. 4(i), reveals their similar character to that discussed above. The two new parabolic bands are localized in the two detached QLs and strongly hybridized with each other, while the two M-shaped bands are localized within the expanded vdW gaps. However, the parabolic band with the bottom at $E_{B}$ = 410 meV observed in ARPES (Fig. 2(d)) was not seen in the present calculation. We tentatively ascribe this discrepancy to a larger expansion of vdW gap than 36\%, which may produce a similar trivial state in the bulk energy gap (see the case of expansion by 5$\rm \AA$ in Supplementary Information). In this case the Dirac state relocates deeper, under second QL (Fig. 4(i)), but its spin structure remains unchanged.

To summarize, we have established that the deposition of Ag on the surface of Bi$_{2}$Se$_{3}$ leads to an intercalation of Ag atoms into the van der Waals gaps between the quintuple layers. This is manifested in forming triangular shaped convexities in the STM image of the surface. As a result, the van der Waals gap expands, and the topological surface state relocates beneath the outermost QL. Our $\it ab initio$ calculations show that this is a consequence of the expanded vdW gap.The detachment of the outermost QL leads to the emergence of trivial surface states, with the states split off the conduction band being localized within the detached QL, and those appearing in the projected gap of the valence band localized on the borders of the enlarged vdW gap. The new findings have major implications for the construction of spintronics devices, where the material is exposed to ambient conditions and the surface becomes contaminated: the buried topological state of the Ag intercalated crystal would be protected from the scattering by adsorbates at the surface. Furthermore, the novel discovery of the topological states existing at the interface created by intercalation process is of great potential to multiply the contribution of the topological transport by constructing large amount of topological channels.

\begin{acknowledgments}
We thank Shuichi Murakami for fruitful discussions. The measurements were performed with the approval of the Proposal Assessing Committee of HSRC (Proposal No.10-A-32, 11-B-40). This work was partly supported by KAKENHI (20340092), Grant-in-Aid for Scientific Research (B) of Japan Society for the Promotion of Science. Author MY thanks financial support from JSPS Research Fellowship. We acknowledge partial support from the University of the Basque Country (Grant No. GIC07IT36607) and the Spanish Ministerio de Ciencia e Innovaci\'on (Grant No. FIS2010-19609-C02-00). Calculations were performed on SKIF-Cyberia supercomputer of Tomsk State University and Arina supercomputer of the University of Basque Country.
\end{acknowledgments}

\bibliography{apssamp}

\end{document}